\documentclass[pdflatex,sn-mathphys-num,iicol]{sn-jnl}

\usepackage{graphicx}%
\usepackage{multirow}%
\usepackage{amsmath,amssymb,amsfonts}%
\usepackage{amsthm}%
\usepackage{mathrsfs}%
\usepackage[title]{appendix}%
\usepackage{xcolor}%
\usepackage{textcomp}%
\usepackage{manyfoot}%
\usepackage{booktabs}%
\usepackage{algorithm}%
\usepackage{algorithmicx}%
\usepackage{algpseudocode}%
\usepackage{listings}%

\usepackage{ifthen}
\usepackage{xspace}

\begin{document}

\title{Metastability in the diluted parallel Ising model}

\author*[1,2,3]{\fnm{Franco} \sur{Bagnoli}}\email{franco.bagnoli@unifi.it}

\author[1,2]{\fnm{Tommaso} \sur{Matteuzzi}}\email{tommaso.matteuzzi@unifi.it}
\equalcont{These authors contributed equally to this work.}

\affil*[1]{\orgdiv{Department of Physics and Astronomy}, \orgname{University of Florence}, \orgaddress{\street{via G. Sansone, 1}, \city{Sesto Fiorentino}, \postcode{50019}, \state{(FI)}, \country{Italy}}}

\affil[2]{\orgname{INFN}, \orgdiv{Sect. Florence}}

\affil[3]{\orgdiv{Team Project IMAGES - UMR ESPACE-Dev}, \orgname{University of Perpignan via Domitia UPVD}, \orgaddress{\street{52 Avenue Paul Alduy} \city{Perpignan}, \postcode{66000} \country{France}}}


\abstract{We present some considerations about the parallel implementations of the kinetic (Monte Carlo) version of the Ising model. In some cases the equilibrium distribution of the parallel version does not present the symmetry breaking phenomenon in the low-temperature phase, i.e., the stochastic trajectory originated by the Monte Carlo simulation can jump between the distributions corresponding to both kinds of magnetization, or the lattice can break into two disjoint sublattices, each of which goes into a different asymptotic distribution (phase). In this latter case, by introducing a small asynchronism (dilution), we can have a transition between the homogeneous and the checkerboard phases, with metastable transients.}




\maketitle

\let\at@
\catcode`@=\active
\def@#1{\ifmmode\boldsymbol{#1}\else\at#1\fi}

\let\quot"
\catcode`"=\active
\def"#1"{``#1''}

\renewcommand\P{\mathcal{P}}
\renewcommand\H{\mathcal{H}}
\newcommand\Z{\mathcal{Z}}

\newcommand{\eq}[2][]{%
    \ifthenelse{\equal{#1}{}}{%
        \begin{equation*}
            #2%
        \end{equation*}%
    }{%
        \begin{equation}\label{eq:#1}%
            #2%
        \end{equation}%
    }%
}
\newcommand{\meq}[2][]{%
    \ifthenelse{\equal{#1}{}}{%
        \begin{equation*}%
            \begin{split}%
                #2%
            \end{split}%
        \end{equation*}%
    }{%
        \begin{equation}\label{eq:#1}%
            \begin{split}%
                #2%
            \end{split}%
        \end{equation}%
    }%
}

\newcommand{\Eq}[1]{Eq.~\eqref{eq:#1}}

\section{Introduction}\label{sec:intro}
The well-known Ising-Lenz model is defined as a statistical mechanics model of ferromagnetism~\cite{Ising1925}, and has been applied to many different contexts~\cite{Macy2024}. It can be defined on different lattices or networks, and can be considered to be part of a larger family of the random cluster models, which includes the standard percolation problem and the Potts model~\cite{Potts1952,Wu1982}. It can be extended to disordered couplings, i.e., spin glasses~\cite{Sherrington1975}.

The Ising model can be easily solved (in the sense of computing the partition function and thus any observable) in the one-dimensional regular lattice case~\cite{Ising1925}, on trees (Bethe lattices)~\cite{Jelitto1979,Glasser1983} and for the two-dimensional regular lattice case~\cite{Onsager1944,Yang1952,Glasser1970}. 

The Ising model has also become the classical playground of Monte Carlo simulations~\cite{Newman1999,Binder2001}. In this case one looks for a Markovian  processes whose asymptotic distributions coincides with the equilibrium one. This  process originates an ensemble of stochastic trajectories, which, although having an arbitrary character, can also be studied to investigate their out-of-equilibrium character and equilibrium relaxation properties~\cite{Ma1985}. In particular, this aspect is important for the study of spin glasses which, below a critical temperature, show such a slow relaxation to equilibrium that it is practically never reached~\cite{bernaschi2025microcanonical}. 

The standard approach to Monte Carlo simulations of the Ising model is that of evaluating the acceptance of a single-spin flip in order to obey the detailed balance condition given by the Boltzmann asymptotic distribution. We call this technique ``serial Monte Carlo'' or ``serial Ising model''.  

There have been several studies about speeding up Monte Carlo simulations, and in particular how large portions of the system can be updated in parallel. This is particularly important when using parallel or multi-core computers (or GPU), and/or multi-spin  coding~\cite{Williams1984,Swendsen1987,Newman1999,Block2010-pb,bernaschi2025microcanonical}.

These techniques, illustrated in Section~\ref{sec:parallel}, are carefully designed so to converge to the asymptotic Boltzmann distribution, therefore giving the same results of the serial Monte Carlo.

On the other hand, there have been investigations on 
out-of-equilibrium models given by the application of the serial Monte Carlo updating in parallel to all sites~\cite{Vichniac1984,Derrida1990}. 
This procedure defines a probabilistic cellular automaton, which has in general different properties with respect to the Ising model.

However, for some updating schemes, it is possible to show that this parallel updating scheme still generates the standard Ising probability distribution, as also illustrated in Section~\ref{sec:parallel}.

In the low-temperature phase and for large-enough dimensions (two for the regular lattices), the system undergoes a phase transition and exhibits symmetry-breaking, which is reflected in the serial Monte Carlo simulations by the confinement of the trajectory to one component of the (factorized) probability distribution (ergodicity braking). In other words, this means that, repeating the simulation from a random configuration, one can reach either a positive or negative magnetization phase, depending on fluctuations. 

Using some parallel implementation, however, it is possible to sample at the same time all components, either randomly jumping between the two phases, or having a sublattice that assumes one magnetization and the other the opposite one.

The main point of this paper is that of investigating this latter case, showing that a small coupling between these two sublattices can promote their convergence (or divergence, depending on the coupling). Moreover, there can be instances of metastability (related to a nucleation process) which can result in long transients before switching to the "correct" distribution.

In the standard Ising model this aspect is not very impacting, since the broken symmetry phase is easily identified and therefore one can check which probability distribution is sampling, it may be a problem in disordered systems like spin glasses~\cite{Mezard1986}.  
In these disordered systems, the order parameter is not simply given by an observable like the magnetization, but it is rather defined by the overlap among replicas~\cite{Mezard1986}. In the absence of an external magnetic field, the statistical weight of a configuration or of its opposite is the same. Clearly, if the system decouples into two independent sublattices, which can go in one configuration or its opposite, the overlap between replicas can take quite different values, even for the same parameters (i.e., the same statistical weight).

The scheme of this paper is the following. We recall the definition of the Ising model and its Monte Carlo (kinetic) implementation in Section~\ref{sec:model}, and the equilibrium distribution in case of parallel update in Section~\ref{sec:distributions}. The effects of a partial application of the parallel updating scheme (asynchronism) is shown in Section~\ref{sec:asynchronism}. The relationships with previous works is reported in Section~\ref{sec:discussion}.
Conclusions are drawn in the last section.

This paper is an extended version of Ref.~\cite{BagnoliMetastableStates}.
With respect to the previous versions, we have added the relation with cluster algorithms, the connection with a related model in which the coupling between the two sublattices is applied at the level of the Hamiltonian~\cite{Nareddy2020}, the relationship with nucleation theory~\cite{Rikvold1994} (see Session~\ref{sec:discussion}) and the results of more extended simulations for the scaling of the order parameter with the dilution.

\section{Equilibrium and kinetic Ising model}\label{sec:model}

We quickly recall here the definition of the Ising model. The system is composed by $N$ spin variables $s_i\in \{-1,1\}$, located on the vertices of a graph, defined by an adjacency matrix $a_{ij}\in \{0,1\}$. We shall indicate a configuration of the whole system ($N$ spins) as $@s$. 

The energy of a configuration $@s$ is 
\eq{
    \tilde{\H}(@s)=-\tilde{J} \sum_{ij} a_{ij} s_i s_j,
}
where for simplicity we have assumed zero magnetic field and uniform coupling $\tilde{J}$. 

From now on, we shall consider only regular square lattices, in one or two dimensions. 

We can define a local field 
\eq{
  h_i = \sum_{j} a_{ij} s_j  ,
}
so that 
\eq{
    \tilde{\H}(@s)=-\tilde{J}\sum_{i} h_i s_i. 
}

The equilibrium distribution is 
\eq{
    \P(@s) = \frac{1}{\Z} \exp \bigl(-\beta \tilde{\H}(@s)\bigr),
}
where $\beta$ is the inverse of the temperature and $\Z$ is the partition function. 

We can rescale $J=\beta \tilde{J}$  so that 
\eq{
    \H(@s)=-J\sum_{i} h_i s_i, 
}
and 
\eq{
     \P(@s) = \frac{1}{\Z} \exp\bigl( - \H(@s)\bigr).
}

Given the probability distribution one can in principle compute the expectation value of any observable $A(@s)$,
\eq{
    \langle A \rangle = \sum_{@s} A(@s) \P(@s), 
}
like for instance the magnetization $M$
\eq{ 
    \langle M \rangle = \sum_{@s} \frac{1}{N}\left(\sum_i s_i\right) \P(@s).
}

\subsection{The kinetic Ising model}
Unless one finds a way of explicitly computing the partitions function, the computation of the observables is done by means of Monte Carlo simulations, originating the \emph{kinetic} Ising model.

The idea of Monte Carlo computation is that of building a stochastic trajectory $@s(t)$ so that asymptotically it spends an amount of time on a  configuration proportional to the equilibrium distribution $\P(@s)$, i.e., defining
\eq{
    P(@\sigma, T) = \frac{1}{T} \sum_{t=1}^T [@\sigma=@s(t)],
}
where $[\cdot]$ is a Kronecker delta function which takes value 1 if $\cdot$ is true and zero otherwise, we want to have 
\eq{
    \P(@s)=\lim_{T\rightarrow\infty} P(@s, T).
}

The value of an observable is therefore computed along a trajectory (eventually after an initial transient $T_0$),
\eq{
    \overline{A} = \frac{1}{T} \sum_{t=T_0}^{T+T_0} A(@s(t)).
}

In order to achieve this goal, one builds a Markov process $M_{@s@s'} = M(@s|@s')$ so that $\P(@s)$ is its stationary distribution
\eq{
    \P(@s) = \sum_{@s'} M_{@s@s'}\P(@s').
}

The convergence to the equilibrium  is assured by the detailed balance condition, which implies that, given two configurations $@s$ and $@s'$, the transition probability $M(@s|@s')$ should obey
\eq{
    M(@s|@s')\P(@s') =  M(@s'|@s)\P(@s),
}
i.e., 
\meq[detailed]{  
    \frac{M(@s|@s')}{M(@s'|@s)} &= \exp\Bigl(-\bigl(\H(@s')-\H(@s)\bigr)\Bigr) \\
    &=\exp(-\Delta \H).
}

If the Markov process is irreducible, then the resulting dynamics is ergodic (in the sense that every configuration has nonzero probability of being accessible from every other one) and the asymptotic probability distribution is unique.

There is a large freedom in choosing the configuration $@s'$ given the present one $@s$. Better performances are achieved if the acceptance ratio is about $1/2$, and this is assured if the energy difference between the two configurations $@s$ and $@s'$ is small. The simplest choice is that of flipping just one site, i.e., given a site index $k$, 
\eq{
    \begin{cases}
        s'_i = -s_i & \text{for $i=k$,}\\
        s'_i = s_i  & \text{otherwise.}
    \end{cases}  
}
In practice, a sample follows a random walk with the smallest jump (one lattice site) in the configuration space.

If, as the present case, interactions are local, the computation of the difference in energy only involves  the local field,
\eq{
    \Delta \H = 2 J h_i s_i. 
}

There are two common implementations of the detailed balance condition, the Metropolis-Hastings~\cite{Metropolis1953,Hastings1970} one, for which spin $i$ is flipped with the probability $m_i$
\eq{
    m_i = \min\bigl(1, \exp(-2 J h_i s_i)\bigr),
}
so that the probability of getting $s'_i = 1$ is 
\eq[metropolis]{
    \tau_m(1|h_i, s_i) = \begin{cases} m_i & \text{if $s_i =-1$}\\
    1-m_i & \text{otherwise}
    \end{cases},
}
and the heat bath or Glauber~\cite{Glauber1963} one, that directly gives the probability $\tau_g$ that spin $s'_i$ takes value 1,
\eq[glauber]{
    \tau_g(1|h_i) = \frac{1}{1+\exp(-2 J h_i)}.
}
The Glauber probability is independent of the present value of spin $i$. 

Since the two configurations are identical but for one site, correlations are strong and this causes a bad estimation of fluctuations and variance. Therefore, one performs measurements only at each Monte Carlo step, that is usually assumed to be composed by $N$ individual steps (but better estimations are possible~\cite{Sokal1997,Newman1999}). 

There is no strict rule for choosing the sequence of sites to be updated. Random choices assure smaller correlations, but implies larger computational effort, while a deterministic sequence can be easily parallelized.

\subsection{Parallel Metropolis-Hastings or Glauber updates}\label{sec:parallel}

\begin{figure}[t]
\begin{center}
\begin{tabular}{c}
(a)\\
\includegraphics[width=0.9\columnwidth]{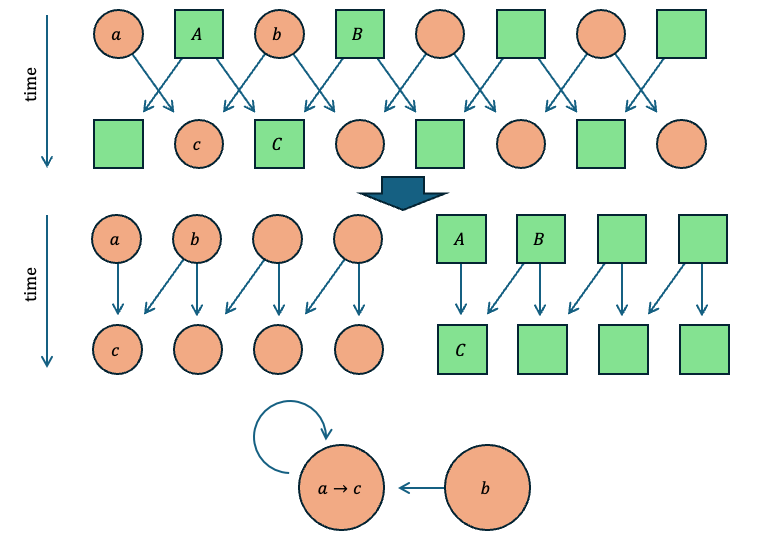}\\
(b)\\
\includegraphics[width=0.7\columnwidth]{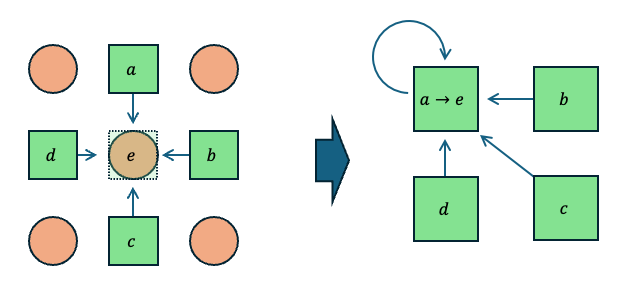}\\
\end{tabular}
\end{center}
\caption{\label{inplace} (a) The separation of the two sublattices for the parallel Glauber dynamics and the in-place update possibility (after having copied the left-boundary spin). (b) Similar procedure for the two-dimensional lattice. }
\end{figure}

In order to fulfill the condition of detailed balance, one can select for the parallel updating all sites whose local field is not affected by the updating procedure. In one dimension, for $N$ even, these are the sites in even positions at even times, whose local field depends only on odd sites, and vice versa (see Fig.~\ref{inplace}-a). In two and three dimensions one can similarly divide the lattice into even and odd sites, computing the parity of the sum of spatial and time indexes (see Fig.~\ref{inplace}-b). 

Explicitly, in $d$ spatial dimensions, let $N = L^d$ ($L$ even) and compute the index 
\eq{
    f = \left(\sum_{k=1}^d i_k + t\right) \mod 2
}
where the $i_k$ denote the index along the $k$-th dimension, and 
\eq{
    i = \sum_{k=1}^d i_k L^{k -1}.
}

The local field of a site with $f=0$ depends only on sites with $f=1$ and vice versa (but the future value of the site has $f=0$). Let us denote the set of sites with same index $f$ at time $t$ as $@s^{(f)}(t)$.

Therefore, for the Metropolis-Hastings algorithm, \Eq{metropolis}, $@s^{(1)}(t+1)$ is computed by using the field computed on $@s^{(1)}(t)$ and the value of $@s^{(0)}(t)$, i.e., the probability that spin $s_i^{(1)}$ takes value 1 is 
\eq{
      \tau_m(s_i^{(1)}=1|h^{(1)}_i, s^{(0)}_i) = \begin{cases} m^{(1)}_i & \text{if $s^{(0)}_i =-1$}\\
    1-m^{(1)}_i & \text{otherwise}
    \end{cases},
}
and vice versa for the sublattice with $f=0$. One can therefore update a whole sublattice in parallel, see for instance Ref.~\cite{bernaschi2025microcanonical}.

For the Glauber algorithm, \Eq{glauber}, the situation is even more extreme. The probability $\tau_g$ that spin $s^{(1)}_i(t+1)$ in the sublattice $f=1$ takes value 1 is the same of \Eq{glauber}, i.e., 
\eq{
    \tau_g\left(s^{(1)}_i=1|h_i^{(1)}\right) = \frac{1}{1 + \exp\left(-2 J h_i^{(1)}\right)},
}
i.e., it depends only on the values of the spins on the same sublattice.

In this latter case, for $L$ even, the system decouples into two non-interacting sublattices, and one can throw away one of them, relabeling the indices, see Fig.~\ref{inplace}. Actually, it is also possible to update the lattice inplace. 

Alternatively, one can apply the same algorithm in parallel to the whole lattice (actually applying them in parallel to the two sublattices), essentially like a probabilistic cellular automaton. We shall use the term "parallel Glauber" for such an algorithm.

\subsection{Swendsen-Wang and Wolff parallel updating scheme}
A different approach to parallel simulation is based on the random cluster model by Fortuin and Kasteleyn~\cite{Fortuin1972}. The random cluster model has brought to the Swendsen-Wang (SW) cluster algorithm~\cite{Swendsen1987}, made  clearer by the Edwards-Sokal representation~\cite{Edwards1988, Grimmett2004}.

The idea of the SW procedure  is that of partitioning the "islands" of same-value spins of the Ising model into statistically correlated clusters, and then flip them with probability $1/2$. The clusters are built by selecting bonds among same-value spins with probability $1-\exp(-2J)$. 

This algorithm quickly samples the equilibrium distribution, even near the phase transition, so that the dynamical critical exponent $z$ is much reduced with respect to the standard single spin-flip algorithm. However, there are cases in which the SW algorithm is not ergodic~\cite{Gore1999}, implying a slow relaxation towards equilibrium, and moreover the construction of small clusters is quite demanding in computational terms. 

A faster algorithm (in the vicinity of the phase transition) is the Wolff's one~\cite{Wolff1989}. In this case one builds just one cluster as in SW case, and flips it. 

These algorithms are quite effective near the phase transition of the homogeneous Ising model, their speed advantage is much less pronounced for spin glasses~\cite{Barzegar2018}. 

\section{Equilibrium distributions of the Wolff and parallel Glauber algorithms}\label{sec:distributions}

\begin{figure}[t]
\begin{center}
\begin{tabular}{c}
(a) \\
\includegraphics[width=0.9\columnwidth]{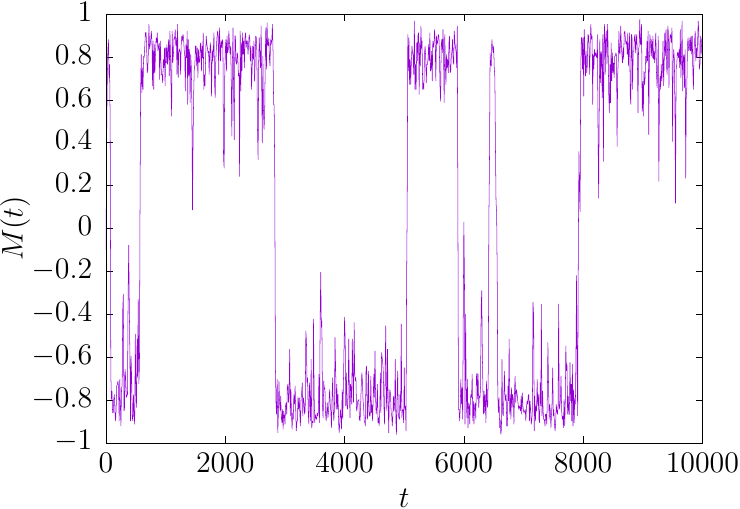} \\
 (b)\\
\includegraphics[width=0.9\columnwidth]{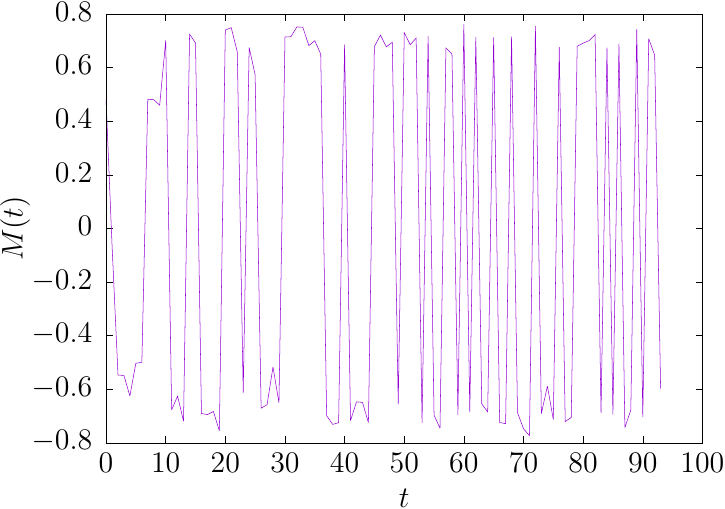}\\
\end{tabular}
\end{center}
\caption{\label{fig:serial-wolff} (a) Time plot of the magnetization $M(t)$ for a $16\times 16$ serial Ising model for $J=0.45$ ($J_c\simeq 0.44$), time measured in Monte-Carlo steps. Due to the small lattice size, magnetization occasionally jumps from positive to negative values, since the asymptotic distribution is not fully factorized. (b) Time plot of the magnetization for a $256\times 256$ Ising model updated using the Wolff algorithm, $J=0.45$, time measured in cluster clips. Despite the large size of the lattice, in this case the magnetization often flips between positive and negative values. Notice that in the sub-figure (b) the time scale is much smaller than in the sub-figure (a).}
\end{figure}

\begin{figure}[t]
\begin{center}
\begin{tabular}{ccc}
(a)&(b)&(c)\\
\includegraphics[width=0.3\columnwidth]{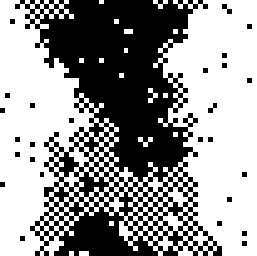}&
\includegraphics[width=0.3\columnwidth]{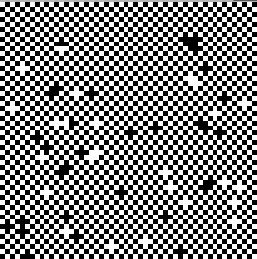}&
\includegraphics[width=0.3\columnwidth]{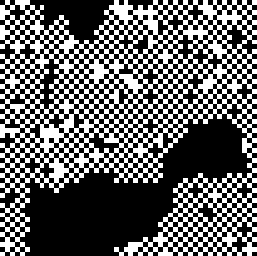}
\end{tabular}
\end{center}
\caption{\label{patterns} Typical patterns in a $50\times50$ lattice (axes $x$ and $y$), for $J=1$ and $t=10000$, where white spots marks negative spins and black spots positive ones. 
(a) The three phases in the fully parallel Ising model ($p=0$), starting from a disordered configuration. Asymptotically, only one phase (black, white or checkerboard) survives. (b) Stability (with fluctuations) of the 
checkerboard pattern (which is the initial state) for small dilutions ($p=0.02$). (c) Droplets growing for larger dilutions ($p=0.045$), again starting from a checkerboard pattern. }
\end{figure}

The Ising model in two dimensions exhibits a ferromagnetic (for $J>J_c\simeq 0.44$) phase transition, with symmetry breaking (with respect to spin inversion) and the factorization of the equilibrium probability distribution. This is reflected by the ergodicity breaking of the related Monte Carlo random walk, in the limit of infinite lattice size. 

This feature can be made evident by examining small lattices: below the critical value $J_c$ of the coupling $J$, the magnetization fluctuates around zero, while for $J>J_c$ (magnetized phase) it takes in average a value grater than zero, with occasional jumps which become rarer with an increasing lattice size, see Fig.~\ref{fig:serial-wolff}-a. 

This is consistent with the fact that the corresponding Markov process, in the limit $N\rightarrow\infty$, is effectively reducible, and the equilibrium distribution factorizes into two distributions, characterized by different average values of the magnetization. 

However, this is not the case for the Wolff algorithm. In the magnetized phase, the cluster (built connecting a fraction of neighboring spins with the same value of the spin) may span most of the lattice and in this case its flipping causes a jump between configurations belonging to the two different equilibrium distributions, see Fig.~\ref{fig:serial-wolff}-b.

On the other hand, in the parallel Glauber algorithm it may happen that, in the broken symmetry phase, one sublattice exhibits positive magnetization and the other a negative one. Since the two lattices exchange their spatial position at each time step, the resulting pattern is a checkerboard (like the ground state of the antiferromagnetic serial Ising model but oscillating in time), see Fig.~\ref{patterns}. 

Notice that a similar checkerboard pattern also arises when applying the Metropolis procedure in parallel to the whole lattice~\cite{Vichniac1984}, but in this case it is due to the simultaneous updating of interacting sites (violation of the detailed balance condition). 

\begin{figure}[t]
\begin{center}
\begin{tabular}{c}
(a)\\
\includegraphics[width=0.9\columnwidth]{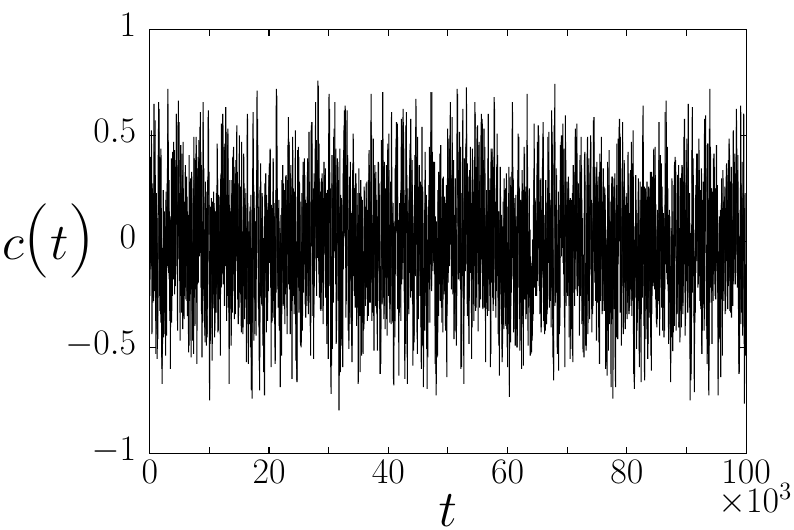}\\
(b)\\
\includegraphics[width=0.9\columnwidth]{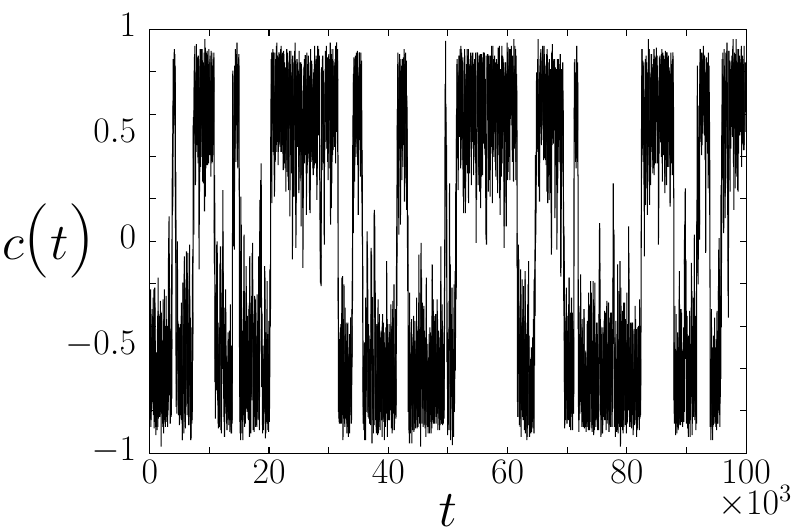}  \\
\end{tabular}
\end{center}
\caption{\label{fig:traj} Time behavior of correlation index $c$ vs time $t$ for   $L=16$ and coupling  (a) $J=0.40$ (below the critical value $J_c\simeq 0.44$, the configuration is disordered and the index $c$ fluctuates around zero), (b) $J=0.45$ (above the critical value $J_c\simeq 0.44$, the system oscillates from homogeneous ($c>0$) and checkerboard ($c<0$) configurations (phases). We use a small value of $L$ ($16$) so that these switch are easily detectable, in the large $N=L\times L$ limit only one phase survives. }
\end{figure}

\begin{figure}[t]
\begin{center}
\includegraphics[width=0.9\columnwidth]{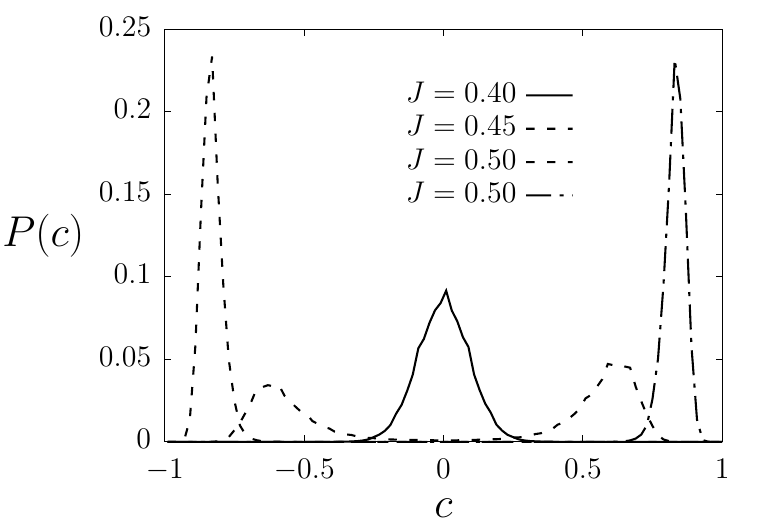}
\end{center}
\caption{\label{fig:distr} Probability distribution of the correlation index $c$ for $L=32$ and several values of $J$. The distribution is computed over one evolution for $2\times 10^6$ time steps after a transient of $10^4$ time steps and using 100 bins. For $J=0.50$ the probability distribution is not unique  (we show two simulations with different asymptotic distributions). The jagged shape of the distribution is due to the binning procedure. }
\end{figure}

\subsection{Correlation index}
In order to easily distinguish between the different configurations (phases), we introduce the correlation index
\eq{
    c(t) = \frac{1}{2dN}\sum s_i(t) h_i(t),
}
where $d$ is the dimensionality of space. 

The correlation index $c$ takes value 1 for the homogeneous (all 1 or all -1) configurations,  value $-1$ for the checkerboard pattern and a value around zero for a disordered pattern . 

The use of the correlation index $c$ allows to easily distinguish between the homogeneous and the checkerboard phases (see Fig.~\ref{fig:traj}).

Starting from a random configuration, one can see in Fig.~\ref{fig:traj}-(a) a typical time evolution of the correlation index $c$ for small values of the coupling $J$ (high temperature) and in Fig.~\ref{fig:traj}-(b) that for  values of the coupling $J$ near the phase transition. In this case, the correlation index $c$ oscillates among  values near  $-1$ and 1, i.e., the configurations are nearly the homogeneous ones ($c=1$) or the checkerboard ones ($c=-1$). 

An alternative view of this behavior is illustrated in Fig~\ref{fig:distr} for $L=32$. In this case we computed the probability distribution of the correlation index $c$ over a single evolution for a large time. For $J=0.40$ the distribution is centered around $c=0$, i.e., disordered configurations dominate. For $J=0.45$ one can see the appearance of peaks near the values $c=\pm 1$, but the distribution is still unique, i.e., these states are metastable, as shown in Fig.~\ref{fig:traj}-(b). For $J=0.5$ the symmetry is broken and the asymptotic distribution is no more unique: repeating the simulation one observes either a peak near $c=1$ or near $c=-1$. 

\begin{figure}[t]
\begin{center}
\begin{tabular}{c}
(a) \\
\includegraphics[width=0.9\columnwidth]{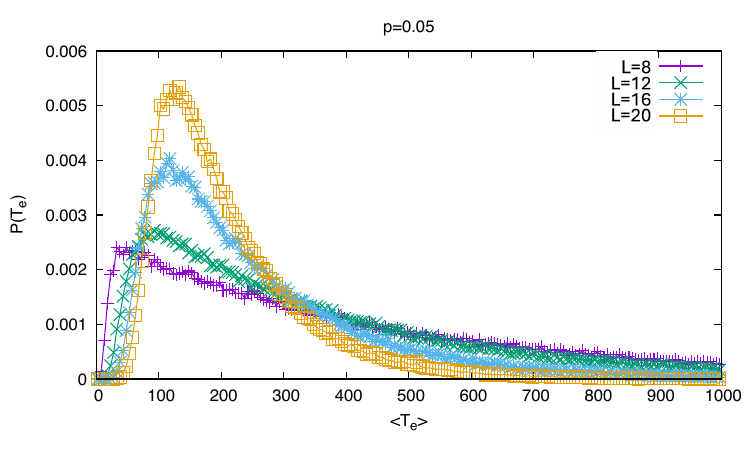} \\
 (b)\\
\includegraphics[width=0.9\columnwidth]{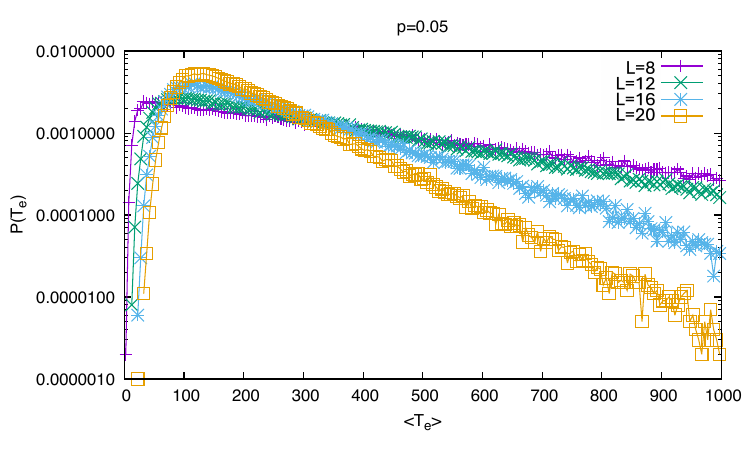}\\
\end{tabular}
\end{center}
\caption{\label{fig:pte} (a) Histogram of the exit time $P(T_e) $ for various values of $L$,  $p=0.0$, $10^5$ samples. (b) The distributions have an exponential tail (here plotted using a logarithmic $y$ scale) from which one can extract a characteristic exit time $\langle T_e \rangle$, which is very near to the average exit time, the difference is given by the non-exponential first part. Error bars are of the order of fluctuations, in general much smaller than symbols. }
\end{figure}

\begin{figure}
    \centering
    \includegraphics[width=0.8\linewidth]{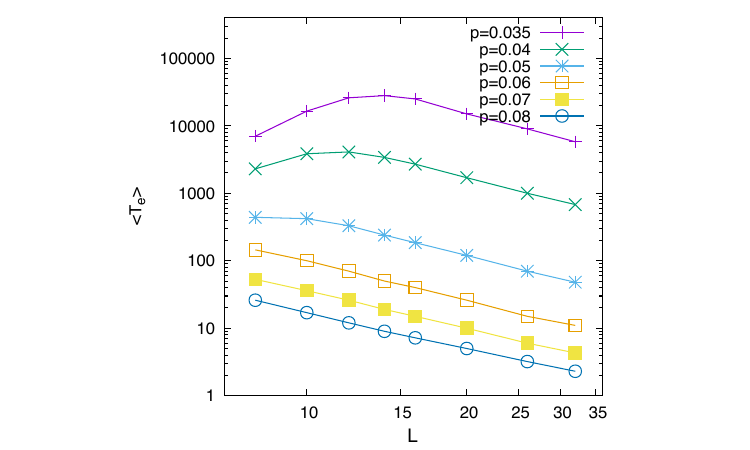}
    \caption{\label{fig:te} The characteristic exit time $\langle T_e \rangle$ for several values of $L$ and of $p$ (logarithmic scale). The dependence of $\langle T_e \rangle$ on $N=L\times L$ seems to follow a common power law (with an exponent roughly $-1.8$), except for small $L$ and $p$. Parameters as in Fig.~\ref{fig:pte}.}
    \label{fig:placeholder} 
\end{figure}

\begin{figure}[t]
\begin{center}
\begin{tabular}{c}
(a) \\
\includegraphics[width=0.9\columnwidth]{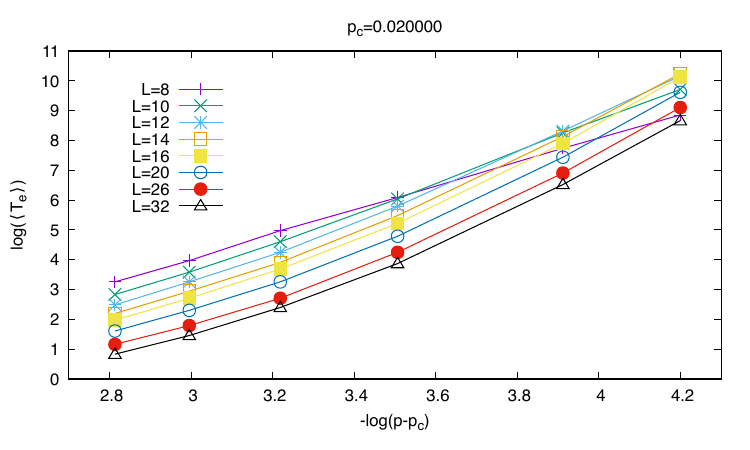} \\
 (b)\\
\includegraphics[width=0.9\columnwidth]{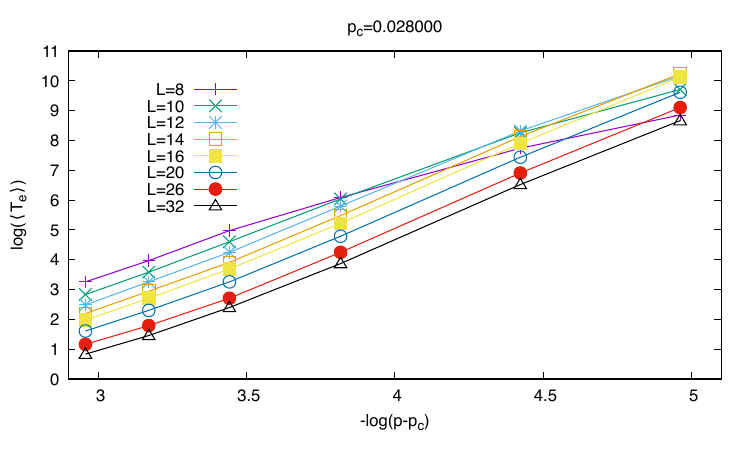}\\
(c)\\
\includegraphics[width=0.9\columnwidth]{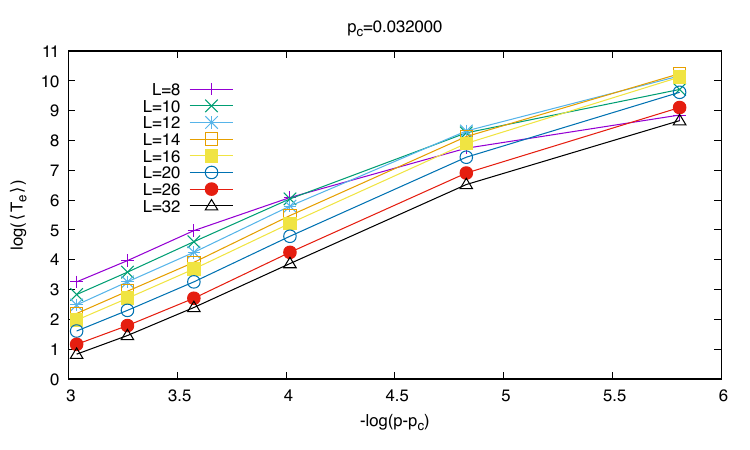}\\
\end{tabular}
\end{center}
\caption{\label{fig:pc} Scaling of the characteristic time $\langle T_e \rangle$ with $p-p_c$ (logarithmic scale) for (a) $p_c =0.02$, (b) $p_c=0.028$ and (c) $p_c = 0.032$. parameters as in Fig. The curves, except those for very small $L$ and $p$, seems to show an asymptotic power-law scaling for $p_c\simeq 0.028$, but it is probably due to finite-size effects. Parameters as in Fig.~\ref{fig:pte}.}
\end{figure}

\section{The effects of asynchronism}\label{sec:asynchronism}

The study of the modifications induced by a partial asynchronism in otherwise parallel systems has been addressed in the context of cellular automata~\cite{Fates2005, Fukś_Fatès_2015}.

Let us now introduce this idea of a partial asynchronism, that we call dilution, for the kinetic Ising model and an  even-$L$ lattices. The control parameter is the fraction $p$ of sites that are not updated, retaining their old values. The dilution couples the two sublattices.

We can extend the transition probability $\tau_g(s'|h)$ including the dilution probability $p$ so that 
\eq[taud]{
\tau_{+}(s'|h, s, p) = 
    \begin{cases}
        [s'=s]&\text{with prob. $p$},\\
            \tau_g(s'|h) & \text{otherwise}.
    \end{cases}
}

The usual serial Ising model corresponds to $p \rightarrow 1$ (say, 1 spin updated per time step, neglecting the null moves), while the fully parallel version to $p = 0$.  

The observables that only depend on single-site properties  take  the same values in parallel or sequential dynamics~\cite{Derrida1990,Neumann1988}.
The presence of the checkerboard pattern however, is due to the strict parallelism of the model.
By diluting the rule, i.e., applying it only to a fraction $p$ of sites, the checkerboard pattern should disappear. This is indeed the case, but for low values of the dilution $p$ the checkerboard pattern is  metastable.

We performed some experiments for $L=8, 12, 14, 16, 20, 26, 32$ and $J=1$ (well inside the magnetized phase). We started with a checkerboard configuration, and updated it with a small dilution $p$. We monitored the correlation index $c$ and defined the escape time $T_e$ from the checkerboard ``basin'' as the time after which $c$ becomes greater than zero (starting from its value $c=-1$ in the checkerboard state). 
This procedure is similar to the inversion of the field in the study of nucleation in the standard Ising model, see Section~\ref{sec:discussion}.

Notice that one can also force the appearance of the checkerboard pattern by applying the dilution with inversion, starting from the homogeneous phase, i.e.
\eq[nottaud]{
\tau_{-}(s'|h, s, d) = \begin{cases}
				[s'=-s] &\text{with probab. $p$},\\
				\tau_g(s'|h) & \text{otherwise}.
				\end{cases}
}

This behavior is qualitatively similar to that induced by a direct coupling between sublattices, see Section~\ref{sec:discussion}.

As shown in Fig.~\ref{fig:pte}, the distribution of the escaping time has an exponential tail 
\eq{
P(T_e)\simeq A \exp\left(-\frac{t}{\langle T_e\rangle}\right),
}
and therefore the standard deviation and the average escape time are very near to the  characteristic time $\langle T_e \rangle$ (the discrepancy is given by the initial non-exponential part). 

We have computed the characteristic exit time $\langle T_e \rangle$ by fitting the exponential tail of the distributions, and, as shown in Fig.~\ref{fig:te}, it shows a power-law dependence with respect to $N$, except for small values of $N$ and $p$.

As shown in Fig.~\ref{fig:pc}, the numerical analysis seems to indicate a critical value $p_c\simeq 0.028$, probably due to finite-size effects. Indeed, one can see that smaller lattices show more pronounced deviations from the power law. However, it is not easy to numerically improve the investigation, since for larger lattices, the variation of the dilution parameter $p$ induces large variations of the characteristic time $\langle T_e\rangle$ that  quickly  passes from extremely large values (long simulations) to such small values (of the order of some units) that the fitting is impossible. 

This finite-size behavior is similar to that of bootstrap percolation~\cite{Adler1991}, where larger lattices apparently show smaller critical threshold.

\section{Discussion}\label{sec:discussion}

The dilution mechanism (parameter $p$), induces a coupling between the two sublattices. A somewhat similar coupling can be introduced directly in the Hamiltonian, as done in Ref.~\cite{Nareddy2020}. In this case the model is defined by the local probability transitions
\eq{
    \tau(s'|h, s) = \frac{\exp((Jh+Ks)s')}{2\cosh(Jh+Ks)},
}
which directly couple the two sublattices (the local field $h$ only depends on the values of spins in one sublattice, while the previous spin value $s$ comes from the other sublattice).

Our dilution is like an annealed discrete realization of this coupling. Instead of coupling all spins, we choose a fraction $p$ of them and replace their value with that of the other sub lattice.

Following Ref.~\cite{Cirillo2001}, it is possible to determine the couplings in the Hamiltonian corresponding to the transition probabilities, which correspond to the detailed balance condition 
\meq{
    \frac{M(@s|@s')}{M(@s'|@s)}&=\frac{\prod_i \tau(s'_i|h_i, s_i) }{\prod_i \tau(s_i|h'_i, s'_i) } \\
    &=\prod_i \frac{\cosh(Jh'_i+Ks'_i)}{\cosh(Jh_i+Ks_i)},
}
where the nominator simplifies because in the exponential there are only products of the kind $s_is'_j$, which factorize.  

Interpreting this expression as in \Eq{detailed}, and exploiting the fact that the spins $s$ are essentially Boolean variables (so that $f(s)=f(1) (1+s)/2 + f(-1) (1-s)/2$), one can obtain the explicit expression of the coupling terms, which connect spins in the neighborhood (and therefore on a given sublattice) and with the other sublattice~\cite{Nareddy2020}.

Unfortunately, this approach cannot be followed for our dilution approach, since the formulation of the transition probabilities of \Eq{taud} cannot be expressed as exponential functions (except maybe in average).

However, numerical experiments (not reported here) show that a qualitatively similar behavior occurs: a positive coupling $K$ destabilizes the checkerboard phase and favors the homogeneous ones, as our ``identical'' dilution of \Eq{taud}, while a negative value of $K$ favors the checkerboard phase~\cite{Nareddy2020}, as our ``opposite'' dilution of \Eq{nottaud}. 

We can also view the effects of the coupling as a kind of  nucleation phenomena~\cite{McDonald1962,McDonald1963,Abraham2012-fl,Burton1977}, see also Fig.~\ref{patterns}.

Our derivation of the exit time $\langle T_e\rangle$ is analogous to the phase change following the inversion of an external magnetic field (or, equivalently, of spins in a constant magnetic field) as done in Ref.~\cite{Rikvold1994}. 

It is also well known that in systems with short-range interactions metastable states eventually decay, even though their lifetimes may be extremely long.~\cite{Rikvold1994}.

We  observed that, for a fixed dilution parameter $p$, there exist a critical size of a droplet such that smaller droplets are reabsorbed and larger ones grow. This observation explains why the characteristic time $\langle T_e\rangle$ is larger for smaller lattices (Fig.~\ref{fig:te}): the probability of a random formation of a droplet of a given size diminishes with the lattice size.

The application of a magnetic field in a standard Ising model is like coupling all spins to a "ghost" one, which never changes its value. Here, we have two  coupled subsystems, so the value of the "ghost field" is not constant. However, what happens in the magnetized phase is that, due to fluctuations, one of the two subsystems suddenly abandons its metastable state and passes to the other, by means of growing ``bubbles'' of the opposite phase. This transition is generally so fast that it is unlikely that the other subsystem initiates the opposite process, at least locally, so that one can approximate the "other" subsystem as static, i.e., a magnetic field.

But in reality, in our set-up also the ``ghost system'' fluctuates, and this may be the origin of the coexistence of nucleation (i.e., a ``first order'' phase transition) and the power-law scaling of the exit time (which should be related to a fluctuation-induced  -- ``second order'' transition).

\section{Conclusions}

We have presented some aspects related to the transition from the classical, serial Ising model to the parallel versions. In the case of parallel updating using the Wolff algorithm, the resulting trajectory samples the whole (factorized) probability distribution also in the symmetry-broken phase (i.e., we do not have ergodicity breaking). Using a parallel Glauber dynamics, one may have a complete decoupling between sublattices, each of which can sample a different probability distribution, giving origin to a checkerboard pattern (similar to the antiferromagnetic phase, but oscillating in time). 

We have shown that 
the checkerboard  patterns are unstable with respect to the dilution of the updating, but that this transition exhibits  finite-size effects and long-lasting metastable states. 

We have also shown that the coupling between sublattices induced by dilution is qualitatively similar to a direct coupling between sublattices in the transition probabilities or in the Hamiltonian~\cite{Nareddy2020}. 

We have also shown that there is a striking analogy between dilution and the presence of an external field, so that the observed metastable states can be considered corresponding to the instability of the unfavorable phase in correspondence of a magnetic field. 

The results of this investigation can be useful in cases in which the broken symmetry phase is not easily recognizable, like in spin glasses~\cite{Mezard1986}. In such systems, metastable states are extremely common~\cite{Cieplak1987,Biroli2001, Abalmasov2023}, and it is important to distinguish those due to the ultrametric structure of phase space to those due to the asynchronism. Indeed, in a parallel implementation of a disordered Ising model, it may happen that sublattices go in different minima of the free energy, but they are not easily recognizable as in the homogeneous systems. Small levels of asynchronism (that may be commonly present in physical systems) would induce metastability and therefore sudden transitions, which may be mistaken for other relaxation processes.  

There are still several open fields of investigation, like mapping the dilution to an effective coupling among subsystems, and clarifying the analogies between metastable states and nucleation.

\subsubsection*{Acknowledgements} 
We are much grateful to Jon Machta for fruitful discussions. 

\section*{Declarations}

\begin{description}
\item{\textbf{Funding:}} This publication was produced with the co-funding of European Union - Next Generation EU, in the context of The National Recovery and Resilience Plan, Investment 1.5 Ecosystems of Innovation, Project Tuscany Health Ecosystem (THE), CUP: B83C22003920001.
\item{\textbf{Conflict of interest/Competing interests:}} Not Applicable
\item{\textbf{Ethics approval and consent to participate:}} Not Applicable
\item{\textbf{Consent for publication:}} Not Applicable
\item{\textbf{Data availability:}} Not Applicable
\item{\textbf{Materials availability:}} Not Applicable
\item{\textbf{Code availability:}} The simulation code is available upon request from the authors. 
\item{\textbf{Author contribution:}} All authors contributed equally. 
\end{description}

\bibliography{ising}

\end{document}